\documentclass[11pt]{article}
\usepackage{moriond,epsfig}

\def\Journal#1#2#3#4{{#1} {\bf #2}, #3 (#4)}

\def\NIMA{{\em Nucl. Instrum. Methods} A}
\def\NPB{{\em Nucl. Phys.} B}
\def\PLB{{\em Phys. Lett.}  B}
\def\PRL{\em Phys. Rev. Lett.}
\def\PRD{{\em Phys. Rev.} D}

\def\be{\begin{equation}}
\def\ee{\end{equation}}
\def\bea{\begin{eqnarray}}
\def\eea{\end{eqnarray}}

\begin{document}
\vspace*{4cm}
\title{Belle CP violation in $b \to s q \bar{q}$ and $u \bar{u} d$ processes}

\author{ C.C.Wang }

\address{Department of Physics, National Taiwan University, \\
No. 1, Sec. 4, Roosevelt Rd, Taipei, Taiwan, 106, R.O.C.}

\maketitle
\abstracts{
We report the analysis of time-dependent $CP$ violation in neutral B meson
system using a data sample corresponding to 140fb$^{-1}$ collected by Belle
detector and KEKB $e^+e^-$ collider. One B meson decaying into $CP$ eigenstate
is fully reconstructed and the accompanying B meson flavor is identified
by its decay products. The $CP$ violation parameters are determined from the
distribution of proper time intervals between the two B decays. Here we
cover measurements of $CP$ violation in  $b \to u \bar{u} d$ and
$b \to s q \bar{q}$ processes.
}

\section{Introduction}
In the standard model (SM) of elementary particles, $CP$ violation
arises from the Kobayashi-Maskawa (KM) phases in weak 
interaction quark-mixing matrix.\cite{ckm} In particular, the SM predicts $CP$
asymmetries in the time-dependent rates for $B^0$ and $\overline{B^0}$
decays to a common $CP$ eigenstate.\cite{tda_pdf} We report the 
time dependent studies of $B^0\to \pi^+ \pi^-$, $B^0\to \phi K_S$,
$B^0\to K^+K^-K^0_S$, and $B^0\to \eta' K^0_S$. The $B^0\to \pi^+ \pi^-$
decay is sensitive to the $CP$-violation parameter $\phi_2$ which is 
dominated by $b\to u\bar{u}d$ transition. Direct $CP$ violation may
also occur in this decay because of interference between $b\to u$ 
tree ($T$) and $b\to d$ penguin ($P$) amplitudes.\cite{PT} For the charmless
decays of $B^0\to \phi K_S$,
$B^0\to K^+K^-K^0_S$, and $B^0\to \eta' K^0_S$ which is mediated by
$b\to s\bar{s}s$, $s\bar{u}u$, and $s\bar{d}d$ transitions are sensitive to
new $CP$-violation phases from physics beyond the SM. \cite{bsm}  

\section{Time-dependent $CP$ violation at Belle}
The Belle detector is a large-solid-angle general 
purpose spectrometer that consists of a silicon vertex detector (SVD),
a central drift chamber (CDC), an array of aerogel threshold 
\v{C}renkov counters (ACC), time-of-flight scintillation counter (TOF),
and an electromagnetic calorimeter comprised of CsI(Tl) crystals (ECL)
located inside a superconducting solenoid coil that provides 1.5 T 
magnetic field.\cite{Belle}  An iron flux return located outside the coil is 
instrumented to detect $K_L^0$ mesons and identify muons.

The Belle detector collects the data at the KEKB asymmetric-energy
$e^+e^-$ collider, in which 8.0 GeV $e^-$ collides with
3.5 GeV $e^+$ at the $\Upsilon(4S)$ resonance.\cite{KEKB} 
The $\Upsilon(4S)$ is produced with a Lorentz boost of 
$\beta\gamma$ = 0.425 nearly along the electron beamline ($z$). 
In the decay chain $\Upsilon(4S) \to B^0\overline{B^0} \to f_{CP} f_{\rm tag}$,
where one of the $B$ mesons at time 
$t_{CP}$ decays to the final state $f_{CP}$
and the accompanying $B$ decays at time $t_{\rm tag}$ decays to the final state 
$f_{\rm tag}$ which distinguishes between $B^0$ and $\overline{B^0}$. The decay 
rate has a time dependence given by 
\be
\label{eq:dt}
\mathcal{P}(\Delta t) = {e^{-|\Delta t|/\tau_{B^0}}\over4\tau_{B^0}}
\{1+q \cdot [\mathcal{S}{\rm sin}(\Delta m_d\Delta t)+\mathcal{A}cos(\Delta m_d\Delta t)]\},
\ee
where $\tau_{B^0}$ is the $B^0$ lifetime, $\Delta m_d$ is the mass difference
between the two $B^0$ mass eigenstates, $\Delta t = t_{CP} - t_{\rm tag}$, and
the $b$-flavor charge $q$ = +1 (-1) when accompanying $B$ meson is a $B^0$ 
($\overline{B^0}$).\cite{tda_pdf}  $\mathcal{S}$ and $\mathcal{A}$ are mixing-induced 
and direct $CP$-violation parameters, respectively.
Since 
$B^0$ and $\overline{B^0}$ are approximately at rest in the $\Upsilon(4S)$
center-of-mass system (cms), $\Delta t$ can be determined from 
the displacement of $z$ between the  
two $B$ mesons: 
$\Delta t\simeq (z_{CP}-z_{\rm tag})/\beta \gamma c\equiv \Delta z/\beta \gamma c$.

\section{Time-dependent $CP$-violation Analysis}
\subsection{Event Extraction}
For reconstruction of $B^0\to \pi^+ \pi^-$ candidates, we use 
the oppositely 
charged track pairs which are identified as pions to form the
$B$ meson. 
The reconstruction of $B^0\to \phi K_S$,
$B^0\to K^+K^-K^0_S$, and $B^0\to \eta' K^0_S$ events from the following
intermediate meson decay chains: $\eta'\to\rho^0(\to \pi^+\pi^-) \gamma$
or $\eta' \to \pi^+\pi^-\eta(\to \gamma\gamma)$, $K_S\to \pi^+\pi^-$, and
$\phi\to K^+K^-$. Candidate $K_S\to \pi^+\pi^-$ and $\phi\to K^+ K^-$ decays
are selected with the same criteria as those used for the previous branching
fraction measurement.\cite{phik} The $K^+K^-$ pairs are rejected  if they
are consistent with  $D^0\to K^+K^-$, $\chi_{c0} \to K^+K^-$, or 
$J/\psi \to K^+K^-$ decays. $D^+ \to K^0_S K^+$ candidates are also removed.
We also applied the same selection criteria for $B^0\to \eta' K^0_S$ 
as our previous analysis.\cite{etapks}
The pion and kaon identification are 
according to the combined information from ACC 
and the CDC $dE/dx$ measurements .

Candidate $B$ mesons are reconstructed using the energy difference
$\Delta E \equiv E^{\rm cms}_{B} - E^{\rm cms}_{\rm beam}$ and the beam-constraint 
mass $M_{\rm bc}\equiv \sqrt{(E^{\rm cms}_{\rm beam})^2-(p^{\rm cms}_{B})^2}$, where 
$E^{\rm cms}_{\rm beam}$ is the cms beam energy, and $E^{\rm cms}_{B}$ and
$p^{\rm cms}_{B}$ are the cms energy and momentum of the $B$ candidate.
The signal region for $B^0\to\pi^+\pi^-$ is defined as 
 $|\Delta E|<$ 0.064 GeV, and 
 5.271 GeV/c$^2$ $< M_{\rm bc} <$ 5.287 GeV/c$^2$. 
The definition for 
 $B^0\to \phi K_S$,
$B^0\to K^+K^-K^0_S$, $B^0\to \eta'(\to \rho\gamma) K^0_S$, and
$B^0\to \eta'(\to \pi^+\pi^-\eta) K^0_S$ are $|\Delta E|<$ 0.06 GeV,
$|\Delta E|<$ 0.04 GeV, $|\Delta E|<$ 0.06 GeV, and 
-0.1GeV $<\Delta E<$ 0.08 GeV all with 
5.27 GeV/c$^2$ $< M_{\rm bc} <$ 5.29 GeV/c$^2$.

In order to suppress the $e^+e^-\to q\overline{q}$ continuum background 
($q$ = $u$, $d$, $s$, $c$), we form the signal and background likelihood
function, $\mathcal{L}_s$ and $\mathcal{L}_{BG}$, from the event topology, 
and apply the likelihood ratio selection
on the reconstructed candidates.\cite{pipi}$^,$\cite{sqq}

\subsection{Flavor Tagging}
The flavor of the accompanying $B$ meson is identified from inclusive
properties of particles that are not associated with the reconstructed
$CP$ side decay, the same as Belle ${\rm sin} 2\phi_1$ 
measurement.\cite{phi1} We used two parameters , $q$ and $r$, to 
represent the tagging information. The first, q, is already defined
in Eq.~(\ref{eq:dt}). The parameter r is and event-by-event, MC-determined 
flavor-tagging dilution factor that ranges from $r$ = 0 for no flavor
discrimination to $r$ = 1 for unambiguous flavor assignment. It is used
only to sort data into six $r$ intervals. The wrong tag fraction for the
six $r$ intervals, $w_l$ ($l$ = 1, 6), and difference between $B^0$
and $\overline{B^0}$ decays, $\Delta w_l$ are determined from the data.

\subsection{Vertex Reconstruction}
The decay vertices of $B^0$ meson are reconstructed using tracks
that have enough SVD hits. The vertex position for the $f_{CP}$ decay
is reconstructed using charged tracks, and the tracks from the 
$K_S^0$ decays are excluded. The $f_{\rm tag}$ vertex is determined using charged 
tracks except those used for $f_{CP}$ and tracks that form a $K^0_S$
or a $\Lambda$ candidate. Each vertex position is constrained by the 
interaction point profile with average transverse $B$ meson decay length 
smearing.

\subsection{Event Distribution Function}
We got
373, 106, 361, and 421 events in the signal box 
for $B^0\to \pi^+\pi^-$, $B^0\to \phi K_S$,
$B^0\to K^+K^-K^0_S$, and $B^0\to \eta' K^0_S$ decays respectively.
We determined event distribution function in the $\Delta E-M_{\rm bc}$ plane
for both signal and background to do the signal purity estimation. 

\begin{figure}
\begin{center}
\epsfig{figure=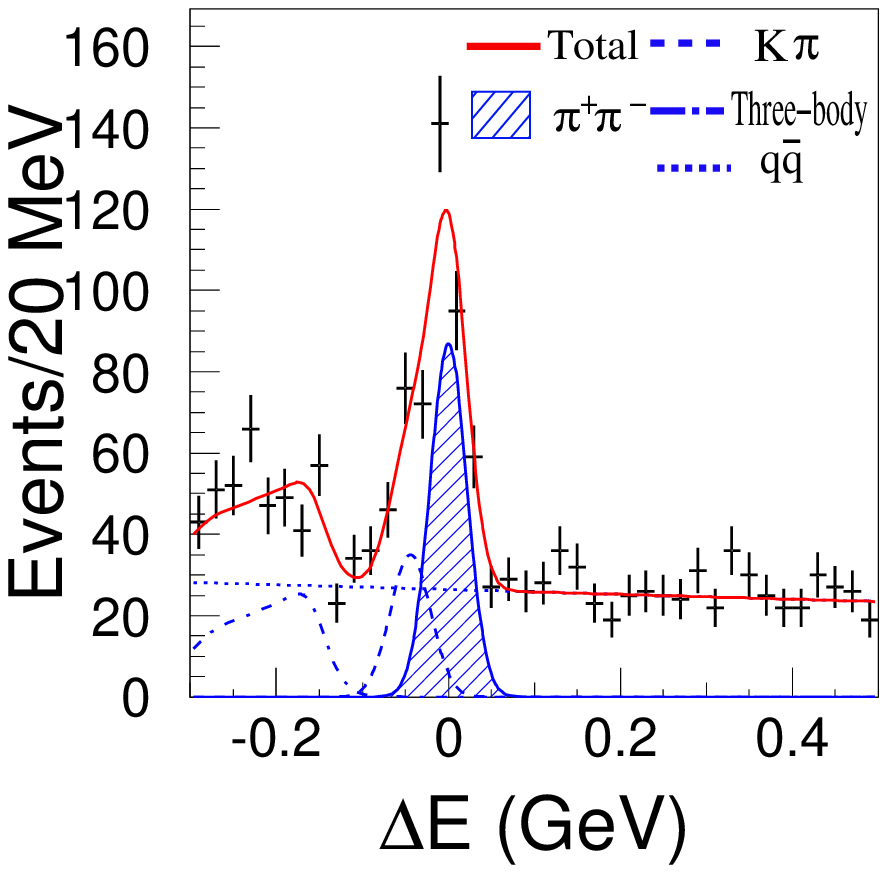,height=3.3in,width=3.1in}
\epsfig{figure=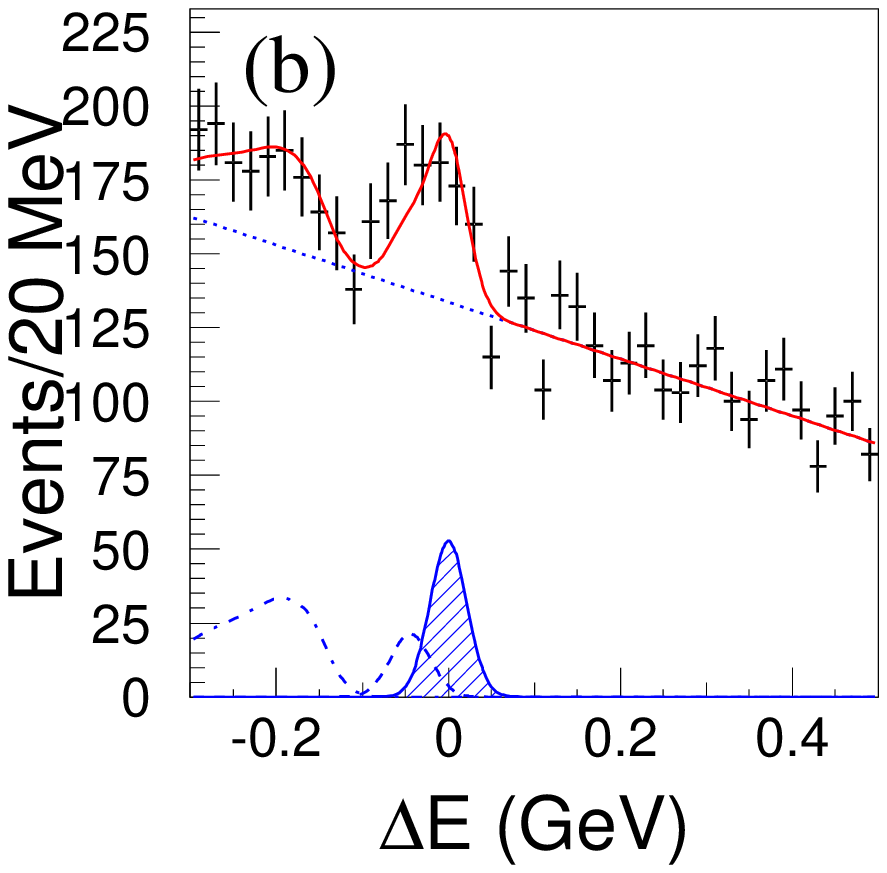,height=3.3in,width=3.1in}
\caption{$\Delta E$ distribution for $M_{\rm bc}$ signal region for 
$B^0\to \pi^+\pi^-$ candidates. The left (right) 
figure is $\Delta E$ distribution
for high (low) purity events. 
\label{fig:pipi}}
\end{center}
\end{figure}

\begin{figure}
\begin{center}
\epsfig{figure=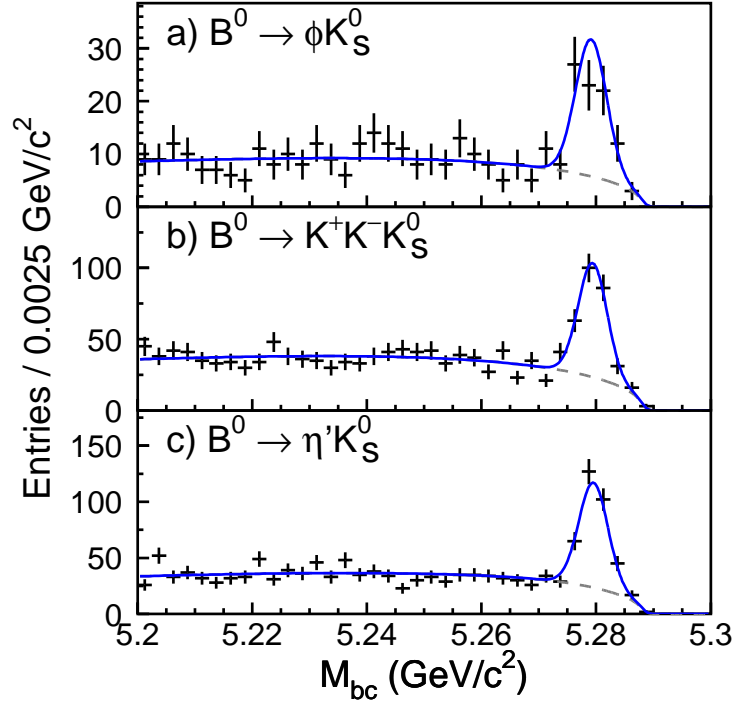,height=4in}
\caption{The beam-energy constrained mass distribution for 
(a) $B^0\to \phi K_S$, (b) $B^0\to K^+K^-K^0_S$, 
and (c) $B^0\to \eta' K^0_S$ within $\Delta E$ signal region. The dashed
curve show the background contributions and the solid curve shows
the fit to signal plus background distribution.
\label{fig:sqq}}
\end{center}
\end{figure}

Figure~\ref{fig:pipi} shows the $\Delta E$ distribution for the 
$B^0\to \pi^+\pi^-$
candidates are in $M_{\rm bc}$ signal region. The background contains 
the continuum $q\bar{q}$, $K\pi$, and charmless three-body $B$ decay
events.
Figure~\ref{fig:sqq} shows the 
$M_{\rm bc}$ distribution for $B^0\to \phi K_S$,
$B^0\to K^+K^-K^0_S$, and $B^0\to \eta' K^0_S$ in $\Delta E$ signal
region. The background is dominated by continuum $q\bar{q}$ events.

\subsection{Maximum-likelihood Fit}
The time-dependent $CP$ violation parameters, 
$\mathcal{S}$ and $\mathcal{A}$, 
are determined by performing an unbinned maximum-likelihood fit. The
probability density function (PDF) expected for the signal distribution is
given by Eq. (\ref{eq:dt}) modified to incorporate the effect of incorrect
flavor assignment. The distribution is convolved with the proper-time 
interval resolution function $R_{\rm sig}$.  A small component of broad 
outliers in the $\Delta z$ distribution is represented by 
$P_{ol}(\Delta t)$. The likelihood for each event is determined
from following PDF:
\bea
\nonumber
 & P_i(\Delta t_i ; \mathcal{S}, \mathcal{A})\\
\nonumber
= & (1-f_{ol})\int^{\infty}_{-\infty}[f_{\rm sig}\mathcal{P}
(\Delta t',q,w_l,\Delta w_l)R_{\rm sig}(\Delta t_i - \Delta t')\\
\nonumber 
+ & (f_{\rm bkg})\mathcal{P}_{\rm bkg} (\Delta t')R_{\rm bkg}(\Delta t_i - \Delta t')]
d(\Delta t')\\
+& f_{ol}P_{ol}(\Delta t_i),
\label{eq:dt_pdf}
\eea
where $f_{ol}$ is the outlier fraction and $f_{\rm sig}$ ($f_{\rm bkg}$) is the 
event-by-event signal (background) 
probability depends on $r$, $\Delta E$ and $M_{\rm bc}$.
$\mathcal{P}_{\rm bkg}(\Delta t)$ and $R_{\rm bkg}$ are the 
background PDF and resolution function. The only free parameters
in the final fit are $\mathcal{S}$ and $\mathcal{A}$, which are 
determined by maximizing the likelihood function
\be
\label{eq:dt_lh}
L = \prod_{i}P_i(\Delta t_i;\mathcal{S},\mathcal{A} ),
\ee
where the product is over all events.

\section{Result}
\subsection{$B^0 \to \pi^+ \pi^-$}
\begin{figure}
\begin{center}
\epsfig{figure=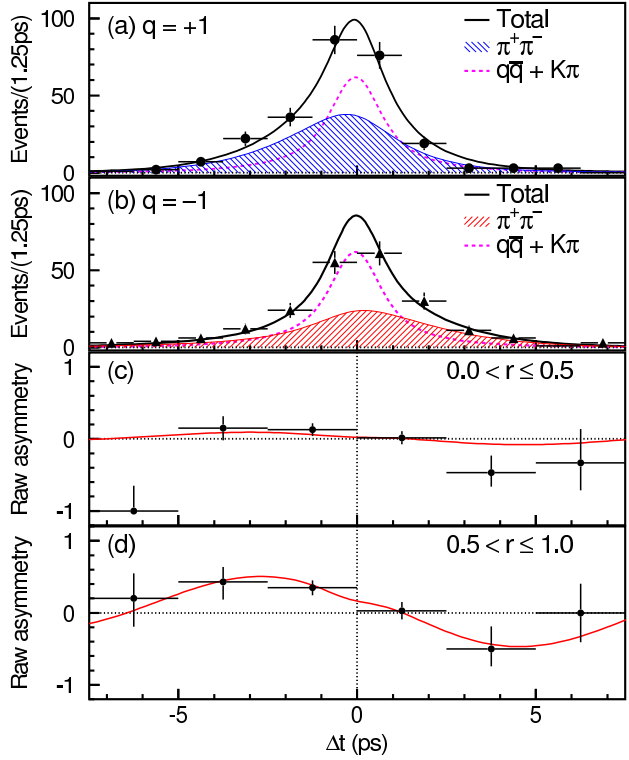,height=4.0in}
\caption{The raw unweighted $\Delta t$ distribution for the 483 
$B^0\to \pi^+\pi^-$ candidates for high purity events in signal region:
(a) 264 candidates with accompany $B$ meson is identified as $B^0$;
(b) 219 candidates with accompany $B$ meson is identified as $\overline{B^0}$. 
(c) and (d) are the raw asymmetry in 0 $<$ $r$ $\le$ 0.5 and 
0.5 $<$ $r$ $\le$ 1 region while the curves show the result of unbinned
maximum-likelihood fit.
\label{fig:pipi_acp}}
\end{center}
\end{figure}

\begin{figure}
\begin{center}
\epsfig{figure=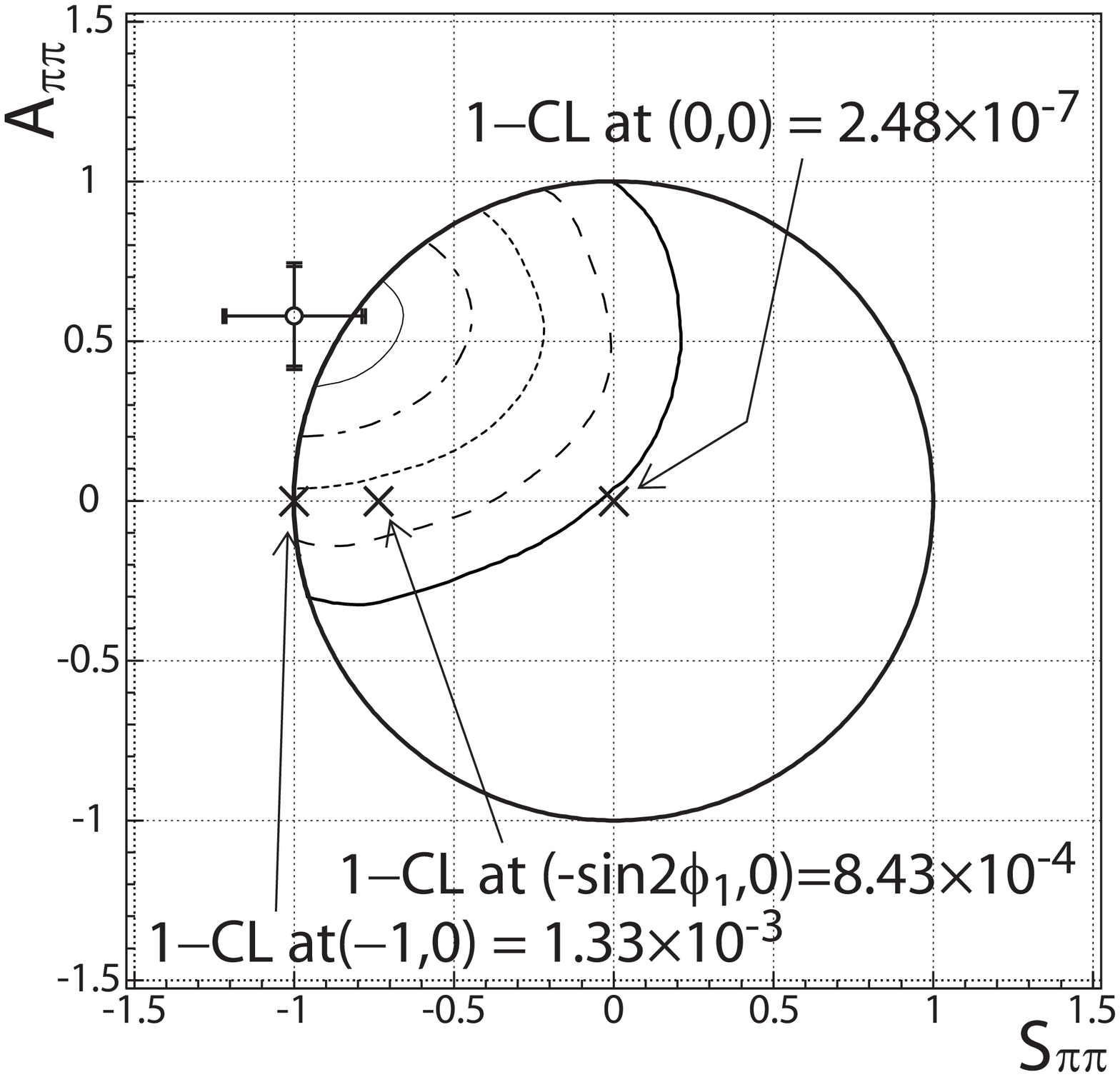,height=4.0in}
\caption{Confidence regions for $\mathcal{S}$ and $\mathcal{A}$. 
\label{fig:pipi_fc}}
\end{center}
\end{figure}

After the unbinned maximum-likelihood fit we got 
$\mathcal{S}$ = $-1.00\pm 0.21(\rm{stat.}) \pm 0.07 ({\rm syst.})$ and
$\mathcal{A}$ = $+0.58\pm 0.15(\rm{stat.}) \pm 0.07 ({\rm syst.})$. 
Figure~\ref{fig:pipi_acp} shows the result. The 
statistical significance is determined from the Feldman-Cousins
frequentist approach.\cite{fc} Figure~\ref{fig:pipi_fc} shows the 
resulting two-dimensional confidence regions in $\mathcal{S}$ and 
$\mathcal{A}$ plane. We have 5.2$\sigma$ significance for $CP$ violation
and 3.2$\sigma$ significance for direct $CP$ violation. If the source
of $CP$ violation is only due to $B-\overline{B}$ mixing or $\Delta B$ 
= 2 transitions so called as super-weak scenarios, 
then the statistical significance for
($\mathcal{S}$,$\mathcal{A}$) = ($-{\rm sin}2\phi_1$,0) is 3.3$\sigma$.\cite{sw1}$^,$\cite{sw2}

The range of $\phi_2$ that corresponds to the 95.5\% confidence level (CL)
for $\mathcal{S}$ and $\mathcal{A}$ in Figure~\ref{fig:pipi_fc} is 
$90^\circ$ $\leq$ $\phi_2$ $\leq$ $146^\circ$ with 
${\rm sin}2\phi_1$ $=$ $0.736$.\cite{phi2_1}$^,$\cite{phi2_2} This result is in agreement 
with constraints on the unitary triangle from other indirect
measurement.\cite{phi2_3} The 95.5\% CL region for 
$\mathcal{S}$ and $\mathcal{A}$ excludes $|P/T|<0.17$.

\subsection{$B^0\to \phi K_S$, $B^0\to K^+K^-K^0_S$, and $B^0\to \eta' K^0_S$}
\begin{figure}
\epsfig{figure=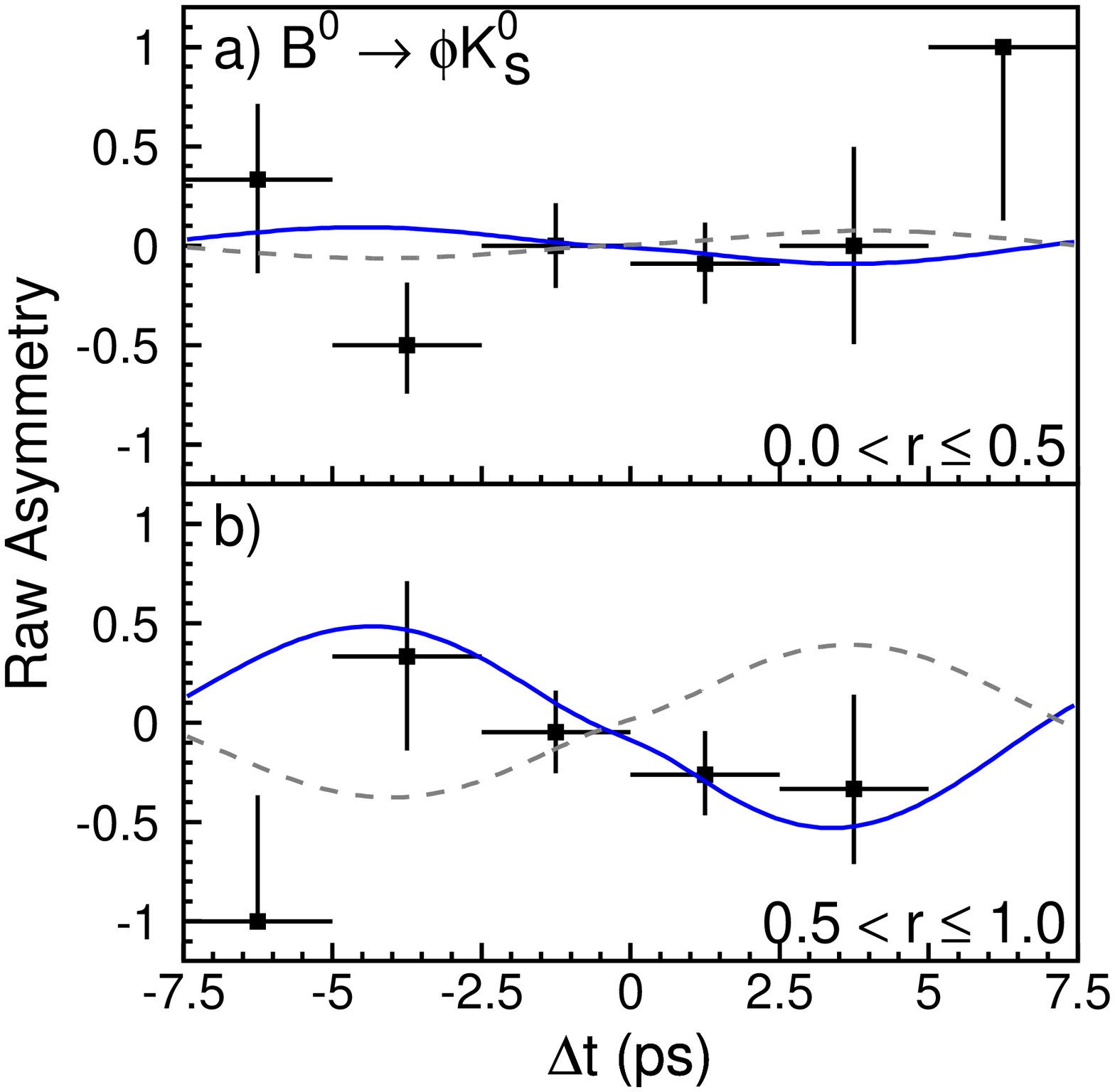,height=2.0in}
\epsfig{figure=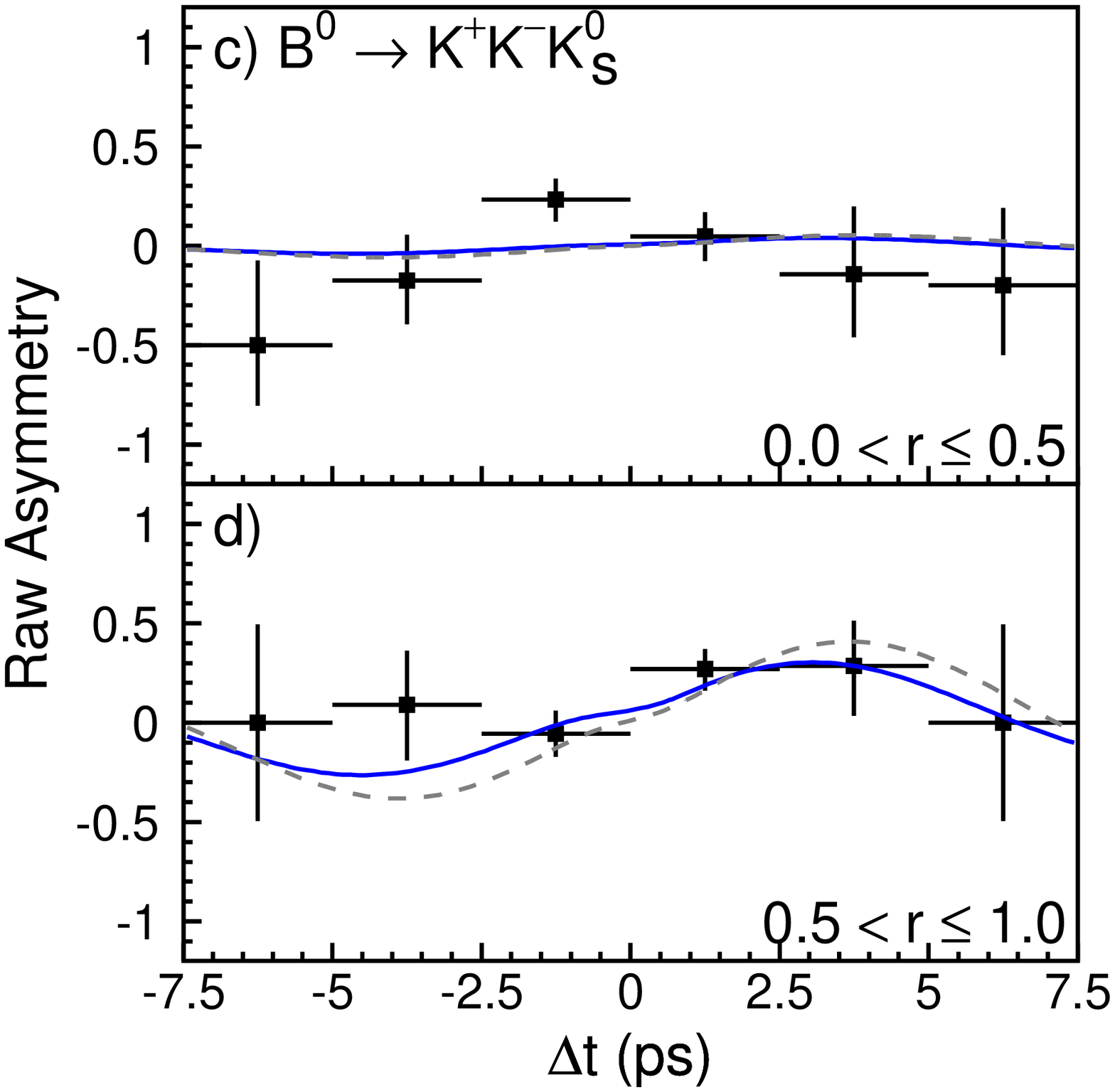,height=2.0in}
\epsfig{figure=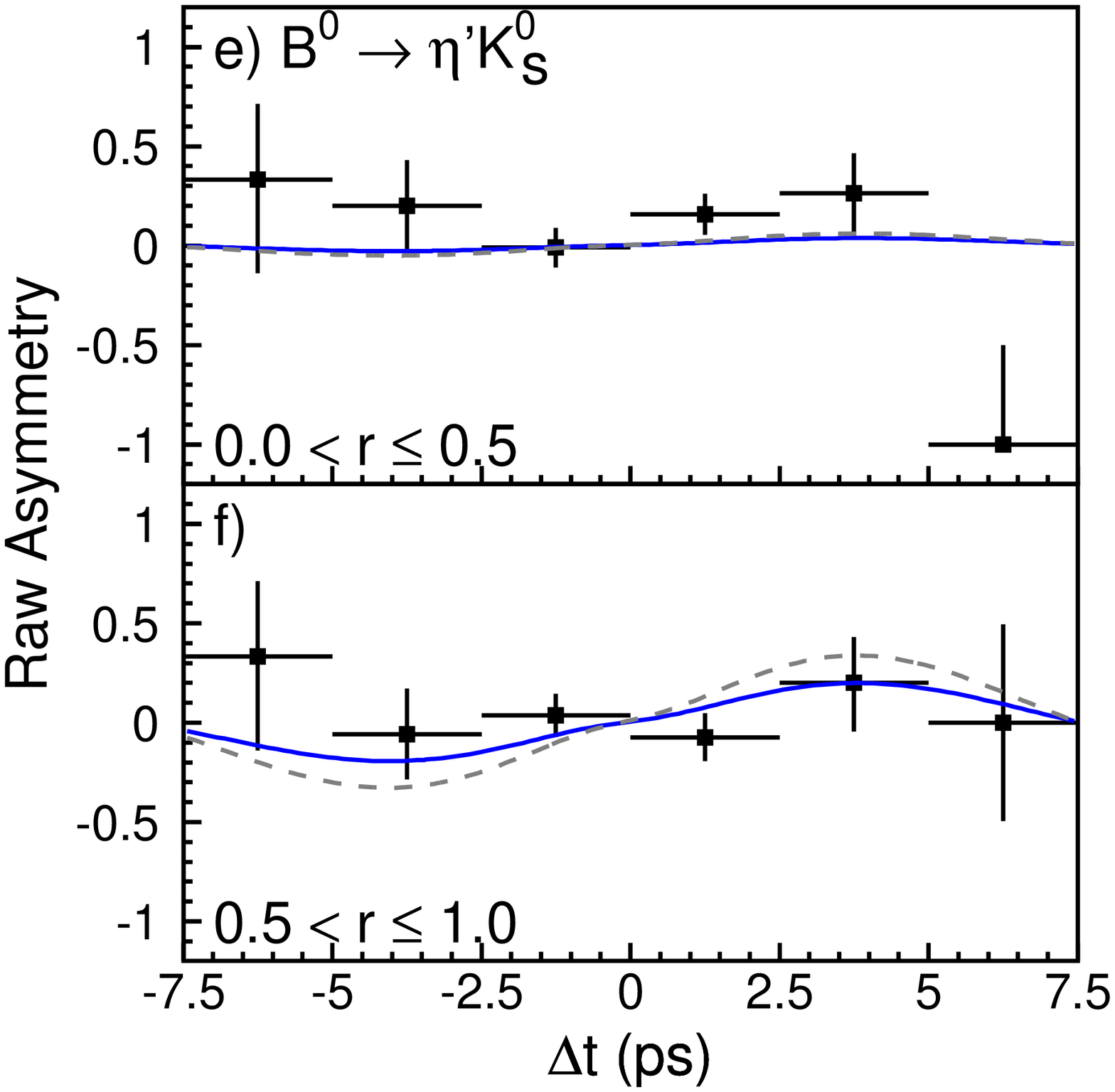,height=2.0in}
\caption{(a)The asymmetry, in each $\Delta t$ bin for $B^0\to \phi K_S$
with 0 $<$ $r$ $\le$ 0.5, (b) with 0.5 $<$ $r$ $\le$ 1.0,
(c) $B^0\to K^+K^-K^0_S$ with 0 $<$ $r$ $\le$ 0.5, (d) with 0.5 $<$ $r$ $\le$ 1.0, 
(e) $B^0\to \eta' K^0_S$ with 0 $<$ $r$ $\le$ 0.5, and (f) with 0.5 $<$ $r$ $\le$ 1.0, 
respectively. In (b) through (g), the solid curves show the result of the unbinned 
maximum-likelihood fit. The dashed curves show the SM expectation with sin$2\phi_1$ = +0.731
and $|\lambda|=1$.
\label{fig:sqqacp}}
\end{figure}

After the unbinned maximum-likelihood fit we got 
$\mathcal{S}$ = $-0.96\pm 0.50(\rm{stat.}) ^{+0.09}_{-0.11} ({\rm syst.})$,
$\mathcal{A}$ = $-0.15\pm 0.29(\rm{stat.}) \pm 0.07 ({\rm syst.})$ for 
$B^0\to \phi K_S$ decay, 
$\mathcal{S}$ = $-0.51\pm 0.26(\rm{stat.}) \pm 0.05({\rm syst.})^{+0.18}_{-0.00}$,
$\mathcal{A}$ = $-0.17\pm 0.16(\rm{stat.}) \pm 0.04 ({\rm syst.})$ for
$B^0\to K^+K^-K^0_S$, and
$\mathcal{S}$ = $+0.43 \pm 0.27(\rm{stat.}) \pm 0.05({\rm syst.})$,
$\mathcal{A}$ = $-0.01\pm 0.16(\rm{stat.}) \pm 0.04 ({\rm syst.})$ for 
$B^0\to \eta' K^0_S$, while the third error for $K^+K^-K^0_S$ mode arises
from the uncertainty in the fraction of the $CP$-odd component.
Figure~\ref{fig:sqqacp} shows the result. Based on Feldman-Cousins
frequentist approach, we get 3.5$\sigma$ statistical significance
of the observed deviation from the SM in $B^0\to \phi K_S$.\cite{fc}

\end{document}